\documentclass[10pt,twocolumn,floatfix]{revtex4}
\usepackage{graphicx}
\usepackage{dcolumn}
\usepackage{bm}
\usepackage{epsfig}
\usepackage{amssymb}
\usepackage{amsfonts}
\usepackage{grffile}
\usepackage[usenames]{color} 
\usepackage{url}
\usepackage{hyperref}
\newif\ifcomment
\def\dvers{v0.65}
\graphicspath{{./pics/}}

\newcommand{\nbin}        {\ensuremath{N_{\rm bin}}}
\newcommand{\aj}          {\ensuremath{A_{\rm J}} }
\newcommand{\pt}          {\ensuremath{p_{\rm T}} }
\newcommand{\pthat}       {\ensuremath{\hat{p}_{\rm T}}}

\newcommand{\GeVc}        {\ensuremath{\mathrm{GeV}/c}}
\newcommand{\misspt}      {\ensuremath{\displaystyle{\not} p_{\mathrm{T}}^{\,\parallel}}}
\newcommand{\missptout}   {\ensuremath{\displaystyle{\not} p_{\mathrm{T}}^{\,\parallel,\,{\rm out}}}}
\newcommand{\vnd}         {\ensuremath{V_{n\Delta}}}
\newcommand{\vtd}         {\ensuremath{V_{3\Delta}}}
\newcommand{\vnj}         {\ensuremath{v_n^{\rm Jet}}}
\newcommand{\vtj}         {\ensuremath{v_3^{\rm Jet}}}
\newcommand{\snn}         {\ensuremath{\sqrt{s_{\rm NN}}}}
\newcommand{\dd}          {\ensuremath{\rm{d}}}
\newcommand{\lsim}        {\,{\buildrel < \over {_\sim}}\,}

\newcommand{\Ref}[1]      {Ref.~\cite{#1}}

\newcommand{\Eq}[1]       {Eq.~\ref{#1}}

\newcommand{\Fig}[1]      {Fig.~\ref{#1}}

\newcommand{\Sect}[1]     {Section~\ref{#1}}

\begin{document}

\title{Remarks on the possible importance of jet $\mathbf v_3$ and multiple jet production \\ 
       for the interpretation of recent jet quenching measurements at the LHC}

\author{C.~Loizides}
\affiliation{Lawrence Berkeley National Laboratory, Berkeley, California, United States}
\author{J.~Putschke}
\affiliation{Wayne State University, Detroit, Michigan, United States}
\date{\today, \color{red}\dvers\color{black}}

\begin{abstract}
Recent jet quenching measurements in Pb+Pb collisions at the LHC report a significant energy 
imbalance of di-jets. The imbalance is found to be compensated by a large amount of soft particles 
produced at large angles with respect to the di-jet axis.
This observation questions the conventional picture of parton energy loss models, established 
at RHIC, which typically expect that the radiated gluons are emitted at moderate angles close to the 
outgoing parton.
In this letter, we qualitatively discuss two possible contributions of the underlying heavy-ion 
background that may have to be taken into account when interpreting the recent data.
We show that a large jet $v_3$, potentially caused by a pathlength dependent energy loss 
in the presence of fluctuating initial conditions, could contribute to the observed excess 
of soft particles apparently originating from large angle in-medium radiation. 
In addition, the observed excess could also be induced by multiple jets produced in the 
vicinity of the leading jet, 
caused by a potential selection bias imposed on the di-jet momentum imbalance.
\end{abstract}


\maketitle
\noindent
 
\section{Introduction\label{sec:intro}}
One of the most important discoveries in the collision of heavy nuclei 
at the Relativistic Heavy Ion Collider (RHIC) is the observation that the inclusive yield
of high transverse momentum ($\pt$) hadrons~\cite{Adcox:2001jp,Adler:2002xw} and the semi-inclusive 
rate of azimuthally back-to-back high-$\pt$ hadron pairs are strongly suppressed relative to the 
expected yields in p+p and d+Au collisions~\cite{Adams:2003im,Adams:2006yt,Adare:2010ry}.
Since high-$\pt$ particles are dominantly produced in the fragmentation of QCD jets, 
the effect has been called ``jet quenching'', expressing the expectation that the parton loses 
energy in the interaction with soft gluons of the medium created in the heavy-ion collisions.

Two different mechanisms have been proposed to describe parton production and evolution 
in the case of a medium:
(a)~collisional energy loss due to elastic scatterings~\cite{Thomas:1991ea,Mustafa:2003vh}, and
(b)~energy loss due medium-induced gluon radiation~\cite{Gyulassy:1993hr,Baier:1996kr,Baier:2000mf}.
Most quantitative attempts to model the jet quenching data from RHIC rely on the radiative energy 
loss scenario as the main dynamical mechanism responsible for the energy loss~\cite{Majumder:2010qh}.

However, as the measurements at RHIC mainly address medium effects concerning the most energetic (leading) 
particle in the jet, the predictions of these models have not really been tested until recently,
when the first measurements of fully reconstructed di-jets in Pb+Pb collisions at $\snn=2.76$ TeV 
at the Large Hadron Collider (LHC) have been reported~\cite{Aad:2010bu,Chatrchyan:2011sx}.
These measurements show a strong increase in the fraction of highly momentum-imbalanced di-jets 
in central as compared to peripheral collisions. 
Furthermore, a large fraction of the momentum imbalance  is found to be carried by a large amount of 
low-$\pt$ ($\lsim 4\GeVc$) particles at large angles ($>45^{\rm o}$) with respect the jet 
axis~\cite{Chatrchyan:2011sx}. 
Whereas the hard component of the momentum distribution of jet constituents reconstructed in Pb+Pb 
collisions is found to be remarkably similar to jets of the same reconstructed energy fragmenting in 
the absence of a medium~\cite{christofrolandqm11}.

These observations are difficult to reconcile with most radiative energy loss models,
established at RHIC, in which the radiated gluons are typically emitted at only moderate angles close 
to the outgoing parton, leading to a characteristic modification of the jet internal 
structure~\cite{Salgado:2003rv,Borghini:2005em}.
The new data from the LHC, qualitatively predicted by \cite{Vitev:2005yg,Vitev:2008rz},
have prompted the development of more sophisticated models~\cite{MehtarTani:2010ma,Qin:2010mn,
CasalderreySolana:2010eh,Young:2011qx,CasalderreySolana:2011rz,Neufeld:2011yh,He:2011pd,Neufeld:2011fh,
Beraudo:2011bh,Chesler:2011nc,CasalderreySolana:2011gx}
to quantitatively describe the di-jet energy imbalance.
It also has been pointed out~\cite{Cacciari:2011tm} that, depending on the 
applied correction scheme, residual contributions of the underlying, soft heavy-ion background and 
its fluctuations can artificially enhance the observed energy imbalance.

In this letter, we discuss the reported excess of soft particles outside of the leading jet cone
of $R=\sqrt{\Delta\eta^2+\Delta\phi^2}>0.8$. 
On the qualitative level we address two possible contributions 
of the heavy-ion background that may affect the current interpretation of the out-of-cone radiation.
The first contribution arises from a potential jet $v_3$, which could be caused by a pathlength 
dependent energy loss in the presence of fluctuating initial conditions~\cite{Alver:2010grk}.
In \Sect{sec:jetv3} we show that under the assumption of a large jet $v_3$ the observed excess 
of soft particles at large angles can be artificially enhanced.
The second contribution arises from multiple, independent, partially overlapping di-jets in the 
heavy-ion event. 
In \Sect{sec:multjets}, for simplicity, we discuss the probability for a second independent 
lower-momentum di-jet pair produced such that one jet is close to the leading jet, while its partner lies 
outside of the cone.
Finally, in \Sect{sec:multjets} and \Sect{sec:aj} we briefly discuss the interplay between the 
mentioned effects, and the observed di-jet energy imbalance $\aj$.

\section{Jet $\mathbf v_3$}
\label{sec:jetv3}

In this section, we qualitatively study the effect of a finite jet $v_3$, on the missing 
momentum~($\misspt$) observable introduced in \Ref{Chatrchyan:2011sx}
\begin{equation} 
  \label{eq:missingpt}
  \misspt = - \sum p_{{\rm T},i} \, \cos \left( \phi_i - \phi_J \right)\,,
\end{equation}
with $\phi_i$ the azimuthal angle of all charged tracks within a certain $\pt$ range 
and $\phi_J$ the azimuthal angle of the leading jet. 
For details concerning jet reconstruction, detector acceptance and kinematical cuts 
we refer to \Ref{Chatrchyan:2011sx}.

\begin{figure}[t!f] 
  \centering 
  \includegraphics[width=0.7\linewidth]{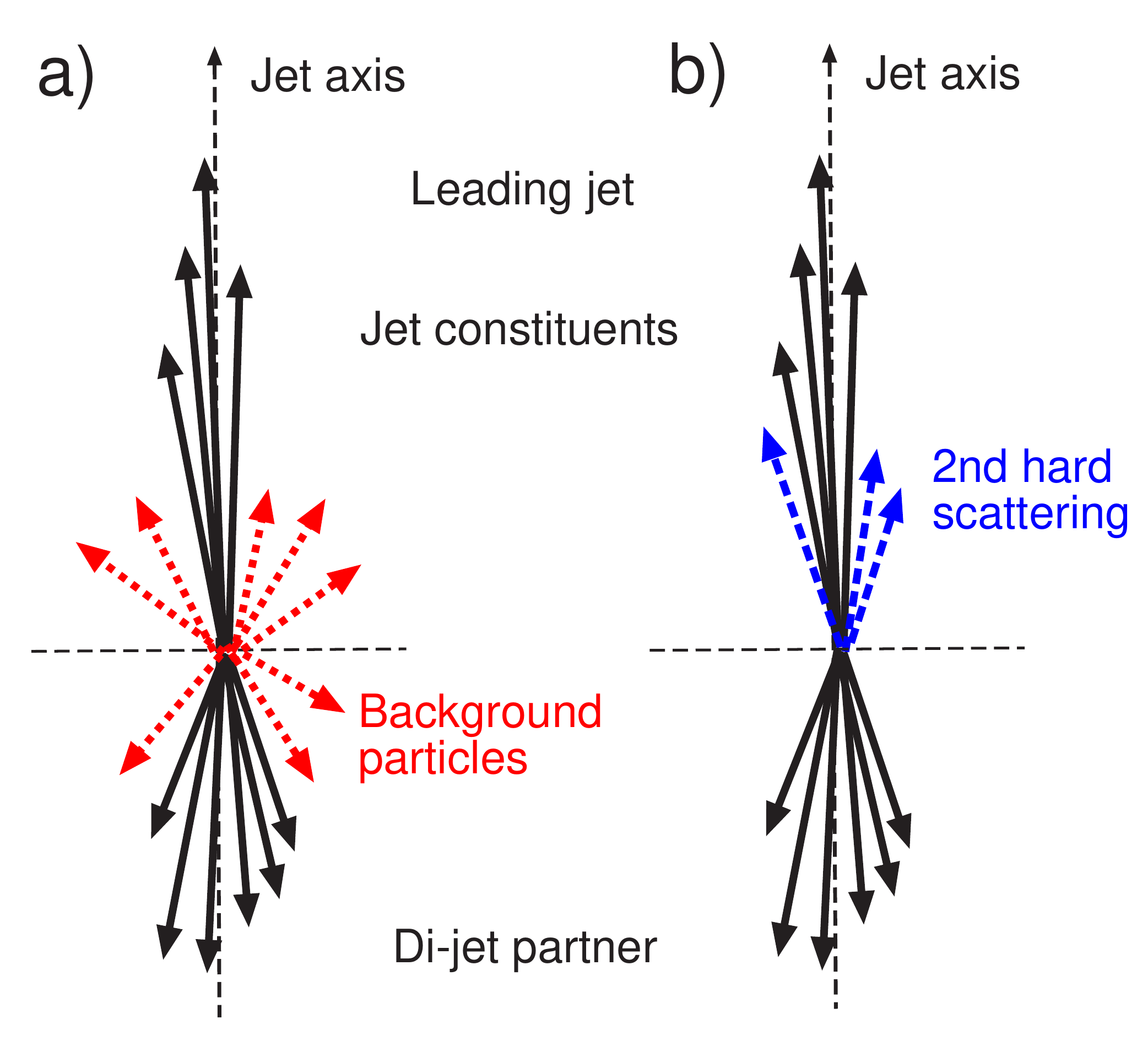}
  \caption{
    \label{fig:jetsketch}(Color online)
    Illustration of the in-cone composition of the particle momentum distribution relative to the jet axis 
    of particles originating from the di-jet and 
    a)~from the $v_3$ background
    b)~from a second hard scattering overlapping with the leading jet, but not with the di-jet partner.
  }
\end{figure}

In \Fig{fig:jetsketch}~a), we illustrate the composition of the particle momentum distribution 
relative to the jet axis of particles originating from the di-jet and from the expected $v_3$ background 
in case of a finite jet $v_3$. 
The main point to note is that the particle distribution caused by $v_3$ is not symmetric around zero 
and $\pi$, as we will outline below.

For convenience, one can express the $\misspt$ observable as an effective two-particle correlation 
using $\Delta\phi= \phi_i - \phi_J$
\begin{eqnarray}
  \label{eq:modelallg}
  \misspt  = - A \int_{\pt^{\rm min}}^{\pt^{\rm max}} \dd\pt \, \pt \, \dd N/\dd\pt 
             \int_{\Delta\phi^{\rm min}}^{\Delta\phi^{\rm max}}  \dd\Delta\phi \nonumber \\ 
             \left(1+2 \sum \vnd(\pt) \cos(n \Delta\phi)\right) \cos(\Delta\phi) \,,
\end{eqnarray}
with $\vnd$ denoting the product of the jet and particle $n^{\rm th}$ order harmonic, 
$\vnd = v_n \, \vnj$ and $\pt \, \dd N/\dd\pt$ the transverse momentum for a given 
kinematic selection of the charged particles in $[p_T^{min},p_T^{max}]$.
Since the single particle harmonics are found to be only weakly dependent on $\eta$~\cite{atlas:2011hfa},
we assume the same for $\vnj$, and so the integration over $\Delta \eta$ has been absorbed into
an overall normalization factor $A$ for the geometrical detector acceptance.

Integration over the full $\Delta\phi$ phase-space ($\Delta\phi^{min}=-\pi$ and $\Delta\phi^{max}=\pi$) 
for all $\pt$ results in $\misspt$ identical to zero for all harmonics, indicating that the soft, flow 
modulated background does not contribute to the $\misspt$ measurement, as also suggested by the 
data~\cite{Chatrchyan:2011sx}. 
This feature makes the $2\pi$-integrated $\misspt$ observable a powerful tool to study jet-quenching 
effects in heavy-ion collisions. 

\begin{figure}[tbt!f]
  \centering 
  \includegraphics[width=0.8\linewidth]{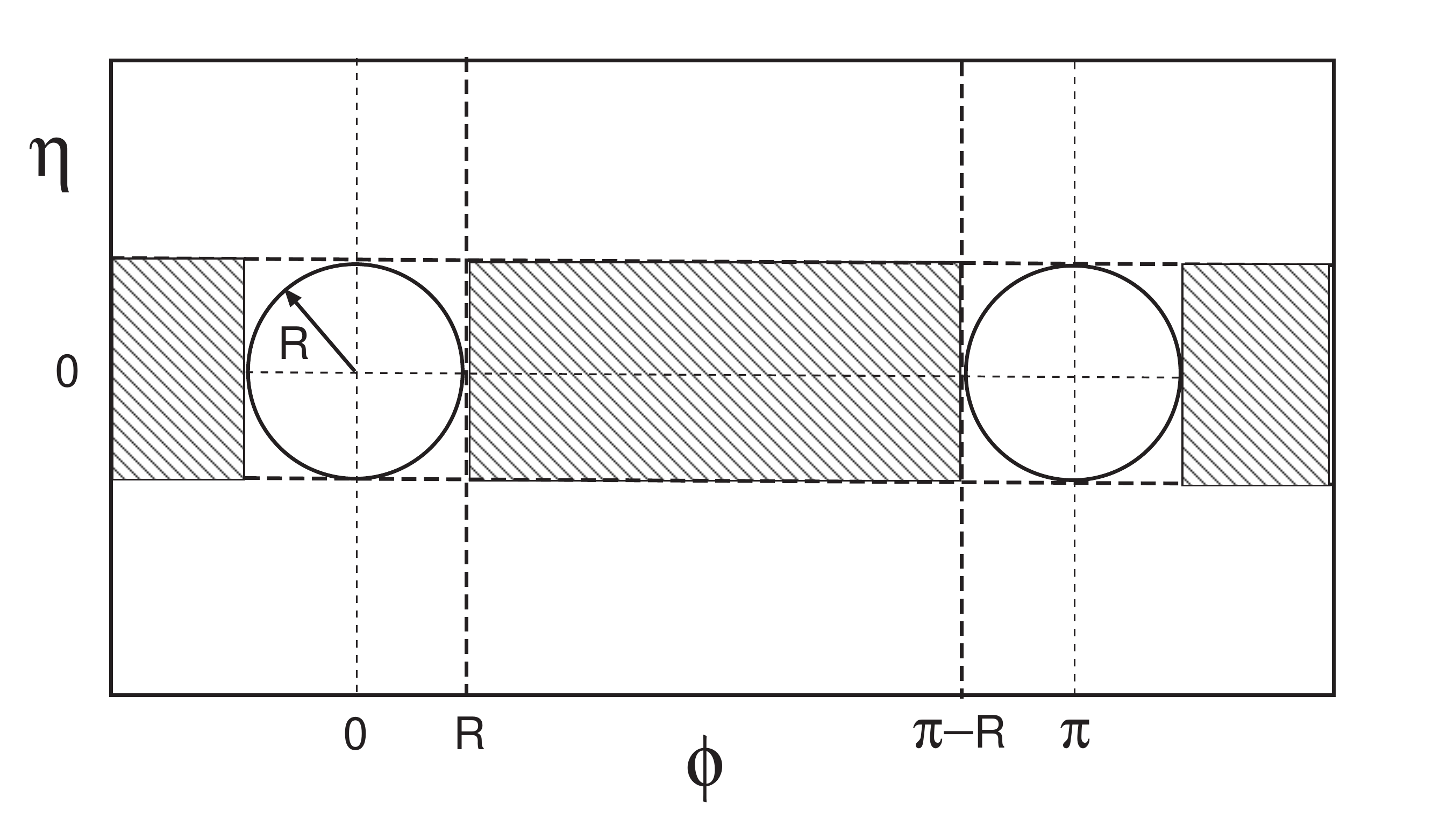}
  \caption{
    \label{fig:modelsketch}
    Simplistic presentation of the considered phase space for \Eq{eq:model}. 
    $R$ is the radius defining the in- and out-of-cone region as used in \Ref{Chatrchyan:2011sx}.
  }
\end{figure}

However, this changes if one considers in- vs.\ out-of-cone contributions to $\misspt$ 
separately from particles in- or outside of the jet cone of radius $R$.
In \Fig{fig:modelsketch} we illustrate the phase space region we use in our simplified, 
but analytical, approach to mimic the exclusion of particles in a jet cone of $R$~($\Delta\phi^{min}=R$ 
and $\Delta\phi^{max}=\pi-R$). 
Since for the $\misspt$ studies the jet selection in \Ref{Chatrchyan:2011sx} has been tightened requiring 
the di-jet partner to be at $\Delta\phi>5/6\pi$ on the away-side, and since we assume no $\eta$ dependence 
of $\vnd$, we can choose the di-jets to be located at ($\eta=0$, $\phi=0$) and ($\eta=0$, $\phi=\pi$).
Using \Eq{eq:modelallg} to estimate a lower limit on the out-of-cone contribution $\missptout$,
we find that odd harmonics do not cancel in $\missptout$, while the even harmonics do.

In the following, we consider only $v_3$ as the expected dominant contribution to $\missptout$
\begin{eqnarray}
\label{eq:model}
  \missptout  = - 4R \int_{\pt^{\rm min}}^{\pt^{\rm max}} \dd\pt \, \pt \, \dd N/\dd\pt 
                \int_{R}^{\pi-R}  \dd\Delta\phi \nonumber \\ 
                \left(1+2 \vtd(\pt) \cos(3 \Delta\phi)\right) \cos(\Delta\phi) \,.
\end{eqnarray}

The missing momentum for particles out-of-cone has to be balanced by the missing momentum for particles 
in the cone. Thus, the fact that the 3$^{\rm rd}$ harmonic is not symmetric around zero and $\pi$, leads to an 
increase of soft background particles in the direction of the leading and consequently to a decrease of soft 
background particles in the direction of the di-jet partner~(see \Fig{fig:jetsketch}). 
This asymmetry results in a decrease of $\misspt$ for soft particles in the direction of the leading jet 
even if the soft quenched particles are fully contained inside the jet cone.
As a consequence this will then result in an enhanced low-$p_T$ contribution in $\missptout$ in 
the direction of the di-jet partner.

\begin{figure}[tbt!f] 
  \centering 
  \includegraphics[width=0.9\linewidth]{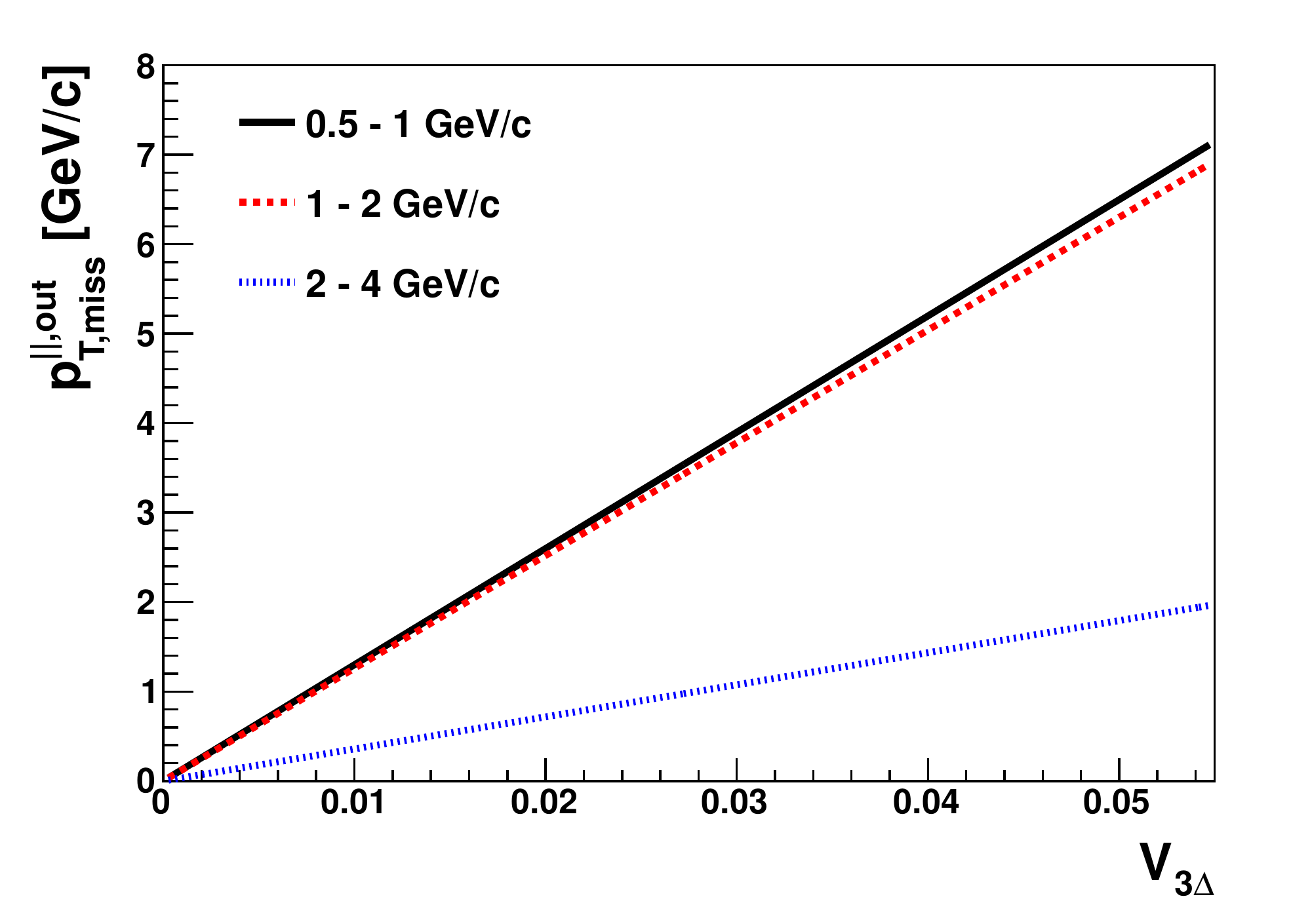}
  \includegraphics[width=0.9\linewidth]{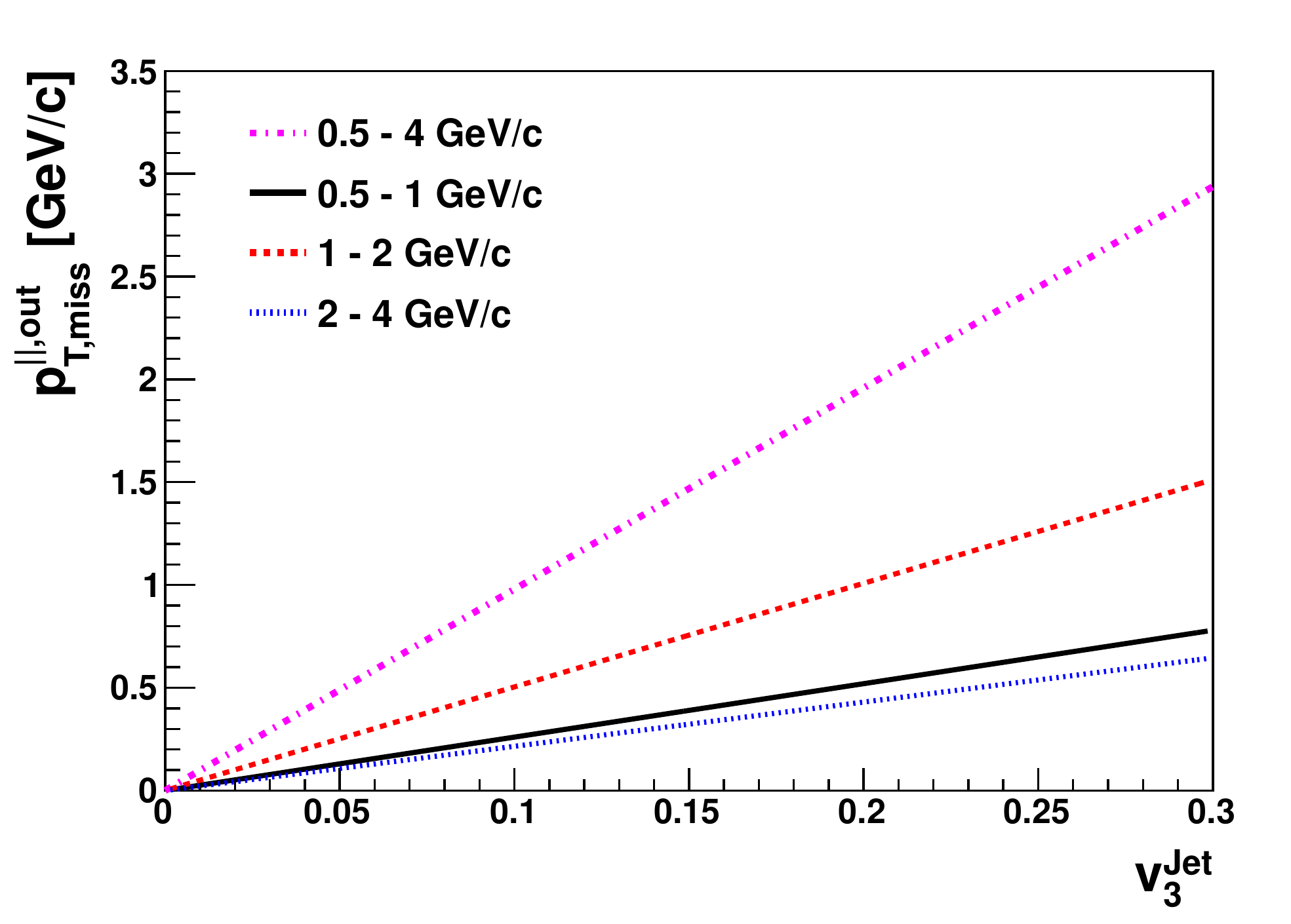}
  \caption{
    \label{fig:v3} (Color online) 
    Missing transverse momentum $\missptout$ (see \Eq{eq:model}) as function of $V_{3\Delta}$ (top)
    and $\vtj$ assuming bulk $v_3$ values from~\cite{Aamodt:2011vk} (bottom)
    for three choices of particle $\pt$ using event multiplicities and transverse 
    momentum spectral shapes for $0$--$30$\% central Pb+Pb collisions.
  }
\end{figure}

In \Fig{fig:v3} (top panel) we show $\missptout$ as function of $V_{3\Delta}$ for three different 
selections of $\pt$ for charged particles with event multiplicities and transverse momentum spectral 
shapes for $0$--$30$\% central Pb+Pb collisions~\cite{Aamodt:2010jd,Aamodt:2010cz}. 
The jet $\vtj$ is unknown, but could be substantial in a path-length dependent energy loss picture, 
and so a finite contribution to $\missptout$ is expected. 
Since $\vtd=v_3 \, \vtj$ one can use the measured bulk $v_3$ values~\cite{Aamodt:2011vk} and with 
assumptions concerning jet $\vtj$, one can estimate the possible contributions to $\missptout$
as illustrated in \Fig{fig:v3} (bottom panel).
For example, assuming a $\vtj=0.2$ the $\pt$ integrated ($0.5$--$4$~$\GeVc$) contribution to 
$\missptout$ can be up to $2$~$\GeVc$.

\section{Multiple jet production}
\label{sec:multjets}

In this section, we qualitatively discuss the effects of multiple jet production. 
The effect of 3-jet events on $\aj$ and $\misspt$, where the third jet is outside of the cone, 
is already taken into account in the p+p reference in \Ref{Chatrchyan:2011sx}, estimated using 
the Pythia Monte Carlo generator
We further compute the 3-jet event probability using NLOJet++~\cite{nlojet}. For similar kinematic 
requirements as in \cite{Chatrchyan:2011sx}, we find the fraction to be less than 10\%. 
(Estimates kindly provided by the authors of \cite{He:2011pd} suggest an even smaller contribution.)
Since 3-jet events are already present in p+p collisions and since we discuss modifications 
with respect to the p+p reference, 3-jet events can therefore not be the only explanation of the 
observed increased di-jet imbalance, nor of the $\misspt$ measurements.

In heavy-ion collisions, however, one expects the cross section for hard processes to scale with 
the number of independent nucleon--nucleon collisions.
Therefore, with respect to a selected di-jet there potentially are up to $\left< \nbin \right>-1$ 
independent additional hard-scatterings above a certain $\pt$  per event, 
which all should be treated as background.
Thus, one can imagine cases where one of the $n^{\rm th}$ hard scatter overlaps with the 
selected leading jet, see \Fig{fig:jetsketch}~b), but its di-jet partner lies outside of
the cone of the selected di-jet system. 

\begin{figure}[t!f]
  \centering 
  \includegraphics[width=0.9\linewidth]{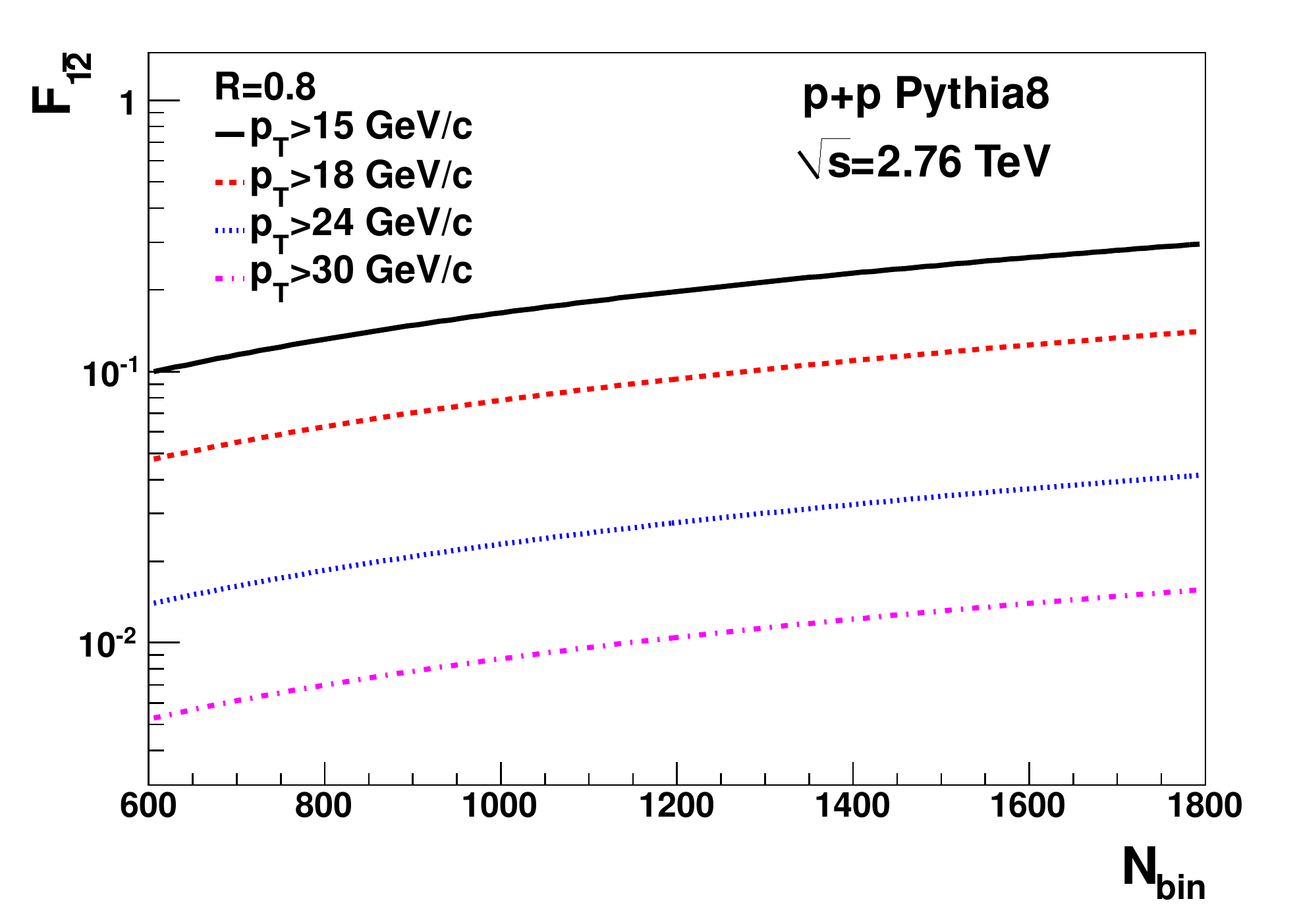}
  \caption{\label{fig:nth} (Color online) 
    Fraction of di-jet events as function of $\nbin$, in which a second hard-scattering overlaps 
    with the leading jet (matching criteria $<R/2$) for different values of minimum $\pthat$.
  }
\end{figure}

Since no background subtraction was performed in the $\misspt$ in- and out-of-cone distributions 
and from the $2\pi$ integrated distribution we know that on average, given the kinematic selection 
and acceptance, the $n^{th}$ hard scattering balance, one has to estimate the fraction of events 
where an additional jet overlaps in $R<0.8$ with the leading jet, but not on the recoil side. 
For simplicity, we estimate the fraction of $2^{\rm nd}$ hard scatterings, $F_{1\overline{2}}$ 
above a certain $\pthat$ threshold using Pythia8~\cite{Sjostrand:2007gs} at $\sqrt{s}=2.76$ TeV 
with a matching/overlap criteria of $R/2$. 
In \Fig{fig:nth}, the fraction for several $\pthat$ selections as a function of $\nbin$ is shown. 
According to our simplistic estimate, the probability of a $2^{\rm nd}$ hard scattering 
is about $8$\% for $\hat{p}_T>18$~$\GeVc$ at $\left<\nbin\right>\approx1000$ (corresponding to 
$0$--$30$\% central collisions) to be in the vicinity of the leading jet and outside the recoil 
cone.
By symmetry, the probability of having the second hard scattering close to the
di-jet partner is equally probable and hence, on average, the two cases should cancel.
However, for larger $\aj$, as discussed in the next section, one can imagine a stronger,
centrality dependent, bias towards events, in which a $2^{\rm nd}$ hard jet is in the 
vicinity of the selected leading jet.
In addition, due to the hard cutoff imposed on the leading jet energy ($>120$ GeV/c in \cite{Chatrchyan:2011sx})
the fraction of these events should be enhanced due to a feed-up of lower energetic jets in the 
measured di-jet selection.

In case the $2^{\rm nd}$ hard jets are quenched, but not fully thermalized, 
they would contribute to the apparent imbalance of the $\misspt$ in-cone, 
especially at lower $\pt$ and consequently to an enhancement in the $\misspt$ out-of-cone 
at lower $\pt$. The effect would cancel if integrated over $2\pi$ in $\misspt$. 
One has to note that only the charged fraction of $\pthat$ in our estimate would contribute to the 
missing $\misspt$ measurement. 
On a qualitative level, the effect of multiple independent jet production will contribute 
to the $\misspt$ in- and out-of-cone measurements and should be taken properly 
into account when interpreting the experimental results.
More quantitative estimates would involve more realistic simulations, addressing the $\aj$ bias, 
and ultimately require assumptions about the underlying partonic energy loss mechanism. 

\section{Effect on $\mathrm \aj$}
\label{sec:aj}
Taking contributions of jet $v_3$ and multiple jet production on the di-jet balance $\aj$ into account, 
the observed, leading jet energy can be expressed as 
\begin{equation}
E_1=\tilde{E}_1+E_{v_3}+E_{nth}\,,
\end{equation}
with $\tilde{E}_1$ the leading jet energy, $E_{v_3}$ the contribution from jet $v_3$ and $E_{nth}$ the 
energy of $n^{\rm th}$ hard scatterings overlapping with the leading jet. 
Similarly the sub-leading, di-jet partner jet energy can be expressed as
\begin{equation}
E_2=\tilde{E}_2-E_{v_3}.
\end{equation}
Therefore, the di-jet imbalance $\aj$ can be written as
\begin{equation}
\aj=\frac{E_1-E_2}{E_1+E_2}=\tilde{A}_{\rm J}+\frac{2E_{v_3}+E_{nth}}{E_1+E_2}\,,
\end{equation}
where $\aj$ is the imbalance of the true di-jet pair~$(\tilde{A}_{\rm J}$) with a positive contribution 
of twice the energy caused by jet $v_3$ and the contribution from multiple hard scatterings. 
For a recent calculation of $\tilde{A}_{\rm J}$ including NLO effects, we refer to \cite{He:2011pd}. 
We would like to point out that estimating the additional effects on $\aj$ without 
a realistic simulation is complicated by the experimentally used background subtraction scheme. 
If the $n^{\rm th}$ hard scatterings in a certain kinematical region, as expected from binary scaling,
are abundantly produced in the experimental phase space, then part of their effect should be accounted 
for in the background correction.

Overall, the discussed effects will lead to an increase in $\aj$, but further, more detailed, 
studies would be needed to quantitatively estimate their contribution on the measured $\aj$ distribution.

\section{Summary}
\label{sec:summary}

Recent jet quenching measurements in Pb+Pb collisions at the LHC report a significant energy 
imbalance of di-jets\cite{Aad:2010bu,Chatrchyan:2011sx}. 
The imbalance is found to be compensated by a large amount of soft particles 
produced at large angles with respect to the di-jet axis~\cite{Chatrchyan:2011sx}.
We qualitatively discuss two effects of the underlying heavy-ion background, 
jet $v_3$ and multiple jet production.

The effect of jet $v_3$ on $\misspt$ out-of-cone in our simplistic approach suggests a moderate, 
but finite contribution of a few GeV/c for realistic bulk $v_3$ values and large jet $v_3$.

The influence of multiple di-jet production per event, 
for $\pthat$ values of the order of the effect~($10$ GeV/c), 
can contribute significantly in the $\misspt$ in- and out-of-cone measurements
provided it is caused by a centrality dependent, bias towards events, 
in which the $2^{\rm nd}$ hard jet is in the vicinity of the selected leading jet.
In addition, due to the hard cutoff imposed on the leading jet energy, 
the fraction of these events should be enhanced due to a feed-up of lower energetic jets in the 
measured di-jet selection.

We want to emphasize that the $\misspt$ measurement over $2\pi$~\cite{Chatrchyan:2011sx}, independent on the 
details of the composition of the leading jet selection and bulk-like correlation effects, is a clear 
indication of partonic energy loss in heavy-ion collisions at the LHC, in which high-$\pt$ suppression 
is balanced by an enhanced production of low-$\pt$ particles. 
However, we think that the discussed effects should be taken into account, and further quantified,
in order to unambiguously conclude whether the quenched energy appears at large angles
with respect to the jet axis.

\section*{Acknowledgments}
We would like to thank Helen Caines for fruitful discussions and critical comments regarding the manuscript.

\bibliographystyle{epj.bst}
\bibliography{jetv3ref}
\end{document}